\documentclass[]{article}

\usepackage{graphicx}
\usepackage{cancel}
\usepackage{authblk}
\usepackage{amsmath}
\usepackage{amsthm}
\usepackage{amssymb}

\usepackage[sort&compress,numbers]{natbib}
\usepackage[colorlinks=true,linkcolor=black, citecolor=blue, urlcolor=blue]{hyperref}

\begin{document}

\title{Self-consistent homogenization approach for polycrystals within second gradient elasticity} 

\author[1]{Y.O. Solyaev}

\affil[1]{Moscow Aviation Institute, Moscow, Russia}

\setcounter{Maxaffil}{0}
\renewcommand\Affilfont{\itshape\small}

\date{\today}

\maketitle

\begin{abstract}
In this paper, we propose a generalized variant of Kr{\"o}ner's self-consistent scheme for evaluation of the effective standard and gradient elastic moduli of polycrystalline materials within Mindlin-Toupin second-gradient elasticity theory. Assuming random orientation of crystallites (grains) we use an extended Eshelby's equivalent inclusion method and mapping conditions between the prescribed linear distribution of macro-strain and corresponding micro-scale field variables averaged over the volume and all possible orientations of single grain. It is found that developed self-consistent scheme predicts the absence of strong gradient effects at the macro-scale level for the model of spherical grains. However, for the more general shape of the grains, considered approach allows to obtain a set of non-linear relations for determination of all effective standard and gradient elastic moduli of polycrystals.
\end{abstract}

\section{Introduction}
\label{int}

One of important application area of Eshelby's equivalent inclusion method is related to determination of the effective elastic constants of polycrystals. By using Eshelby's method, Hershey \cite{hershey1954elasticity} and Kr{\"o}ner \cite{kroner1958berechnung} developed a self-consistent approach for estimation of the effective elastic constants of isotropic polycrystalline materials based on the known elastic properties of randomly oriented anisotropic grains. It was found that this approach provides highly accurate predictions in comparison with experimental data \cite{hershey1954elasticity,kube2015ultrasonic}. Extensions of self-consistent scheme have been proposed for determination of the elastic properties of polycristals with texture \cite{kneer1965berechnung}, cracks \cite{kanaun1980poisson} and reinforcement particles \cite{chen2022multiscale}; also it is used for estimation of the  elastoplastic properties \cite{kroner1961plastischen}, third-order constants within nonlinear elasticity \cite{lubarda1997new}, conductivity properties \cite{sevostianov2019kroner}, attenuation coefficients in ultrasonics \cite{kube2015ultrasonic}, etc. 

In the present study, the generalized variant of Kr{\"o}ner's self-consistent scheme is proposed for determination of the effective material constants of isotropic polycrystalline materials within Mindlin-Toupin strain gradient elasticity theory (SGET) \cite{Mindlin1964,Toupin1964}. Constitutive relations of isotropic SGET involve two standard Lame elastic constants and five additional Mindlin elastic constants \cite{Mindlin1964,dell2009generalized}. Determination of these constants is important for different applications, related e.g. to the refined analysis of size effects in damage evalution and fracture of polycrystalline materials \cite{Askes2015,vasiliev2021new,placidi2022hemi,makvandi2021strain}, in wave propagation processes \cite{nejadsadeghi2020role} and in the multi-scale materials design \cite{dell2019pantographic}. Previously, different homogenization methods have been established within SGET for the composite materials and metamaterials \cite{bacca2013anisotropic,aydin2022strain,solyaev2022self,yang2022verification,rahali2015homogenization,lahbazi2022size,solyaev2022effective}. In the present study we propose a variant of self-consistent homogenisation for polycrystals composed of randomly oriented grains. Assessments on the second gradient moduli of polycrystals via phenomenological grain-pair interaction parameters have been developed previously within SGET in Ref. \cite{barchiesi2021granular}.

In order to found the relations between the effective standard and gradient elastic moduli of isotropic polycrystal and the elastic properties of its anisotropic grains we consider the self-consistent scheme, which implies that each grain is regarded as an inclusion embedded in the homogeneous effective medium represented by the surrounding grains.
We propose to consider the linear distribution of macroscopic strain field (instead of classical homogeneous strain field). Introducing the generalized mapping conditions between the averaged strain and second gradient of displacement inside the single grain and corresponding macroscopic quantities we derive the system of seven non-linear equations for determination of the effective material constants of SGET. Averaged field variables inside the single grain are determined based on the extended Eshelby's equivalent inclusion method that was proposed within SGET for the case of linear distribution of eigenstrain in Refs. \cite{Ma2014,ma2018inclusion, solyaev2022self} 
Based on the obtained results it is shown, that the non-spherical shape of randomly oriented grains can be the source of strong strain gradient effects at the macro-scale level of polycrystals, while for the spherical Cauchy-type grains the method reduces to the classical self-consistent homogenization scheme and trivial solution for the effective gradient moduli.

\section{Second-gradient elasticity theory}
\label{sge}

Consider the formulation of the second-gradient elasticity theory for the macroscopically isotropic linear elastic material occupying a region $V$. The strain energy density within Mindlin Form I is given by \cite{Mindlin1964}:
\begin{equation}
\label{w}
\begin{aligned}
	w(\pmb{\varepsilon}, \pmb\kappa) =
	\tfrac{1}{2}\, \pmb{\varepsilon} : \textbf{C}:\pmb{\varepsilon} +
	\tfrac{1}{2}\, \pmb{\kappa} \,\vdots \,\textbf{G}\, \vdots\, \pmb{\kappa}
\end{aligned}
\end{equation}
where $\pmb{\varepsilon} = \tfrac{1}{2}(\nabla \textbf{u} + (\nabla \textbf{u})^T)$ is an infinitesimal strain tensor,  $\pmb\kappa = \nabla\nabla \textbf{u}$ is the second gradient of displacement; $\textbf{u}(\textbf x)$ is the displacements vector at a point $\textbf x$; $\nabla$ is nabla operator; $:$ and $\vdots$ define double and triple contractions between tensors, respectively; \textbf{C} and \textbf{G} are the fourth-order and sixth-order isotropic tensors, which components are given by \cite{Mindlin1964, ma2018inclusion}:
\begin{equation}
\label{C}
\begin{aligned}
	C_{ijkl} = \lambda\delta_{ij}\delta_{kl} + 2 \mu I_{ijkl}
\end{aligned}
\end{equation}
\begin{equation}
\label{G}
\begin{aligned}
	G_{ijklmn} &= 
	\tfrac{a_1}{2}(I_{lmin}\delta_{jk} + I_{lmjn}\delta_{ik})
	+ \tfrac{a_2}{2}(I_{ijkn}\delta_{lm} + I_{lmkn}\delta_{ij})\\[5pt]
	&+ 2 a_3 \delta_{ij}\delta_{lm}\delta_{kn}
	+ a_4 I_{ijlm}\delta_{kn}
	+ \tfrac{a_5}{2}(I_{ijln}\delta_{km} + I_{ijmn}\delta_{kl}))
\end{aligned}
\end{equation}
where $\lambda$ and $\mu$ are the standard Lame constants; $a_i$ ($i=1...5$) are the additional material constants of gradient theory that can be denoted as Mindlin second-gradient elastic constants or gradient moduli; $\delta_{ik}$ is Kronecker delta, i.e. the second-order identity tensor (which will be denoted as $\textbf I^{(2)}$); $I_{ijkl}=\tfrac{1}{2}(\delta_{ik}\delta_{jl} + \delta_{il}\delta_{jk})$ is the fourth-order isotropic identity tensor (which will be denoted as $\textbf I^{(4)}$) and in the following we will also use the six-order isotropic identity tensor $\textbf I^{(6)}$, which components are given by $(\textbf I^{(6)})_{ijklmn} = I_{ijlm}\delta_{kn}$.

By using standard variational procedure, the following statement of the boundary value problem for the bodies with smooth surface $S$ without edges can be obtained within SGET \cite{Mindlin1964}:
\begin{equation}
\label{BVP}
\begin{cases}
	\nabla \cdot (\pmb{\tau} - \nabla \cdot \pmb{\mu})+ \bar{\pmb b}=0,
	\qquad &\textbf{x}\in V\\
	\textbf{n} \cdot (\pmb{\tau} - \nabla \cdot \pmb{\mu})
	- \nabla_S \cdot(\textbf{n} \cdot \pmb{\mu})
	- H \,\pmb{\mu}_n = \bar{\pmb t},
	\,\,\, or \,\,\,
	\textbf u = \bar{\textbf u}_s,
	 &\textbf{x}\in S\\
	\pmb{\mu}_n = \bar{\pmb m}
	\,\,\, or \,\,\,
	\partial_n \textbf u = \bar{\pmb g},
	\quad &\textbf{x}\in S
\end{cases}
\end{equation}
where the bar symbol denotes the corresponding quantities prescribed in the body volume and on its surface; 
$\pmb\tau = \partial w/\partial \pmb{\varepsilon} =
	\textbf{C}:\pmb{\varepsilon}$ is standard stress tensor; $\pmb{\mu} = \partial w/\partial \pmb \kappa =
	\textbf{G}\,\vdots\, \pmb\kappa$ is the third-order double stress tensor; $\pmb{\mu}_n = (\textbf{n} \otimes\textbf{n}) : \pmb{\mu}$ is the surface double-traction;
	$\nabla_S = \nabla - \textbf{n} \partial_n$ is the surface gradient operator;  $\partial_n$ is the normal gradient; 
	$H = - \nabla_S \cdot \textbf{n}$ is twice the mean curvature of $S$; 
	\textbf{n} is the unit outward normal vector on the boundary $S$.
		
Discussion on the physical meaning of non-classical constants and stresses as well as on the formulation of SGET can be found elsewhere \cite{dell2009generalized,polizzotto2016note}.

\section{Self-consistent second-gradient homogenization for polycrystals}
\label{hom}

Let us consider a polycrystalline material that is an assembly of homogeneous grains with the same anisotropic centrosymmetric elastic properties, defined by constitutive tensors $\textbf C^{gr}$ and $\textbf G^{gr}$ within SGET (the structure of constitutive relations for anisotropic materials within SGET can be found in Refs. \cite{auffray2019complete, po2019green, polizzotto2018anisotropy}). We assume that grains have random crystallographic orientations, i.e. that polycrystal has no texture and it is macroscopically isotropic \cite{kroner1958berechnung, markov2000elementary}. The shapes of grains in a simplest model is approximated by a sphere with mean radius $R$. In more general case, grains shape can be approximated by spheroid or ellipsoid \cite{markov2000elementary}. The overall elastic behaviour of polycrystall is also described within SGET and its constitutive tensors $\textbf C$ and $\textbf G$ are defined by \eqref{C}, \eqref{G}. The goal of homogenization is to derive the relations between the effective constants in tensors $\textbf C$, $\textbf G$ and given properties of grains ($\textbf C^{gr}$, $\textbf G^{gr}$, grains size and shape). 

Following self-consistent approach, we consider single anisotropic grain embedded in the infinite effective isotropic medium with yet unknown  properties $\textbf C$, $\textbf G$. The grain has ellipsoidal shape and the origin of coordinate system is placed at its center. Linear distribution of a strain field is imposed at the macro-scale level so that the displacements, the strain and the second gradient of displacement are given by:
\begin{equation}
\label{qbc}
\begin{aligned}
	\textbf u(\textbf x)  &= 
	\pmb\varepsilon^0\cdot\textbf x
		+ \tfrac{1}{2}\pmb\kappa^0:(\textbf x\otimes\textbf x),\\
	\pmb\varepsilon(\textbf x)  &= \pmb\varepsilon^0 
		+ \hat{\pmb\kappa}^0\cdot\textbf x,\\
		\pmb\kappa(\textbf x)  &= \pmb\kappa= \pmb\kappa^0
\end{aligned}
\end{equation}
where $\pmb\varepsilon^0$ is constant symmetric second-order tensor ($\varepsilon^0_{ij}=\varepsilon^0_{ji}$), which defines the mean strain field at the macro-scale level; and $\pmb\kappa^0$ is constant third-order tensor ($\kappa^0_{ijk}=\kappa^0_{jik}$), which defines the second gradient of displacement at the macro-scale level;  the hat symbol defines symmetrization so that $\hat{\kappa}^0_{ijk}=\tfrac{1}{2}({\kappa}^0_{ijk}+{\kappa}^0_{ikj})$.

Adopted linear variation of strain field \eqref{qbc} can be realized in the Minlind-type equivalent homogeneous material by definition of the so-called quadratic boundary conditions and distributed volume forces (see \cite{Forest2011,monchiet2020strain}). According to Eshelby's equivalent inclusion method generalized for SGET \cite{Ma2014, ma2018inclusion,solyaev2022self}, we can state that under prescribed remote conditions \eqref{qbc} the averaged strain field and averaged second gradient of displacement inside the grain can be defined as follows:
\begin{equation}
\label{esh}
\begin{aligned}
	\langle\pmb\varepsilon^{gr}\rangle_v  
		&= \langle\textbf A\rangle_v : \pmb\varepsilon^0,\\
	\langle\pmb\kappa^{gr}\rangle_v  &= \langle\textbf B\rangle_v \,\vdots\, \pmb\kappa^0
\end{aligned}
\end{equation}
where $v$ denotes the domain occupied by the grain, which volume is $|v|$; $\langle\textbf A\rangle_v$ is the fourth-order strain concentration (localization) tensor averaged over the grain volume; $\langle\textbf B\rangle_v$ is the averaged six-order concentration tensor for the second gradient of displacement averaged over the grain volume; and we will use the  notation with angle brackets for the averaged quantities: 
$$\langle f \rangle_v = \tfrac{1}{|v|}\int_v f(\textbf x) \text{d}v$$

Averaged concentration tensors can be represented based on the Eshelby's equivalent inclusion method as follows \cite{Ma2014, ma2018inclusion,solyaev2022self}:
\begin{equation}
\label{AB}
\begin{aligned}
	\langle\textbf A\rangle_v = 
	\left(\textbf I^{(4)} + \langle\textbf S^{(4)}\rangle_v : (\textbf C)^{-1}
	:(\textbf C^{gr} - \textbf C)\right)^{-1} \\[5pt]
	\langle\textbf B\rangle_v = 
	\left(\textbf I^{(6)} + \langle\textbf S^{(6)}\rangle_v \,\vdots\, (\textbf G)^{-1}
	\,\vdots\,(\textbf G^{gr} - \textbf G)\right)^{-1} 
\end{aligned}
\end{equation}
where $\textbf S^{(4)}$, $\textbf S^{(6)}$ are the position-dependent Eshelby-like tensors that can be derived based on the solutions of inclusion problems within SGET \cite{Ma2014, ma2018inclusion}.

Note, that in classical Kr{\"o}ner's homogenization method the single relation  for the local strain (stress) field inside the grain is involved under prescribed remote homogeneous boundary conditions for the overall strain (stress) state \cite{kroner1958berechnung, hershey1954elasticity, markov2000elementary}. In the present case we propose an extended approach with additional condition for the averaged second gradient of displacement \eqref{esh}$_2$ under prescribed linear distribution of strain. One can also use the similar additional condition for the gradient of strain within Mindlin Form II, while in the present case we use Form I due to known explicit representation of the six-order tensor $\textbf S^{(6)}$ in this theory \cite{ma2018inclusion}.

Let us now average Eq. \eqref{esh} with respect to all possible crystallographic orientations of the grain axes to obtain:
\begin{equation}
\label{av}
\begin{aligned}
	\langle\pmb\varepsilon^{gr}\rangle_{v, \Omega}
		= \langle\textbf A\rangle_{v,\Omega} : \pmb\varepsilon^0,\qquad
	\langle\pmb\kappa^{gr}\rangle_{v,\Omega}
		= \langle\textbf B\rangle_{v,\Omega} \,\vdots\, \pmb\kappa^0
\end{aligned}
\end{equation}
where $\langle...\rangle_{v,\Omega} = \langle\langle...\rangle_{v}\rangle_{\Omega}$ denotes the averaging over the volume and orientations of axes of grain, that is:
$$\langle f \rangle_{\Omega} 
	= \tfrac{1}{8\pi^2}\int_{\Omega}f(\textbf x)  \text{d}\Omega$$
where $d\Omega = \sin\theta \,\text{d}\phi \,\text{d}\theta\, \text{d}\psi$, and $\phi$, $\theta$ and $\psi$ are the Euler angles.

The main idea of the method is the observation that the strain field and the second gradient of displacement averaged over the volume and orientations of grain should be equal to the corresponding prescribed macro-scale fields:
\begin{equation}
\label{cond}
\begin{aligned}
	\langle\pmb\varepsilon^{gr}\rangle_{v,\Omega}
		= \pmb\varepsilon^0,\qquad
	\langle\pmb\kappa^{gr}\rangle_{v,\Omega}
		= \pmb\kappa^0
\end{aligned}
\end{equation}

Combining \eqref{av} and \eqref{cond} we obtain the basic equations of the proposed second-gradient self-consistent method:
\begin{equation}
\label{eqv}
\begin{aligned}
	\langle\textbf A\rangle_{v,\Omega}
		= \textbf I^{(4)},\qquad
	\langle\textbf B\rangle_{v,\Omega}
		= \textbf I^{(6)}
\end{aligned}
\end{equation}

or by using relations for the concentration tensors \eqref{AB}, we have:

\begin{equation}
\label{fin}
\begin{aligned}
\left\langle\left(\textbf I^{(4)} + \langle\textbf S^{(4)}\rangle_v : \textbf C^{-1}
	:(\textbf C^{gr} - \textbf C)\right)^{-1}\right\rangle_{\Omega} = \textbf I^{(4)} 
	\\[15pt]
\left\langle\left(\textbf I^{(6)} + \langle\textbf S^{(6)}\rangle_v\,\vdots\,
\textbf G^{-1}
	\,\vdots\,(\textbf G^{gr} - \textbf G)
\right)^{-1}\right\rangle_{\Omega} = \textbf I^{(6)}
\end{aligned}
\end{equation}

System \eqref{fin} could be reduced to seven coupled non-linear equations for estimation of seven unknown moduli of effective medium $\lambda$, $\mu$ \eqref{C} and $a_i$ ($i=1...5$) \eqref{G} in the following way. Averaging over all directions on the left hand side in relations \eqref{eqv} and \eqref{fin} implies the equality of the linear invariants of the concentration tensors to the corresponding isotropic identity tensors \cite{kroner1958berechnung, lubarda1997new}. Therefore, condition for strain concentration \eqref{eqv}$_1$ reduces to two equations: $$(\langle\textbf A\rangle_{v})_{iijj} = I_{iijj} = 3, \qquad (\langle\textbf A\rangle_{v})_{ijij} = I_{ijij} = 6$$. Conditions for the averaged second gradient of displacement \eqref{eqv}$_2$ results in five equations: 
$$(\langle\textbf B\rangle_{v})_{ijjikk} = I_{ijjikk} = 6, \quad
(\langle\textbf B\rangle_{v})_{iikkjj} = I_{iikkjj} = 3,$$
$$(\langle\textbf B\rangle_{v})_{iikjjk} = I_{iikjjk} = 9,\quad
(\langle\textbf B\rangle_{v})_{ijkijk} = I_{ijkijk} = 18,$$
$$(\langle\textbf B\rangle_{v})_{ijkkji} = I_{ijkkji} = 6$$ 

All these equations are trancendental since the components of averaged Eshelby-like tensors $\langle\textbf S^{(4)}\rangle_{v}$, $\langle\textbf S^{(6)}\rangle_{v}$ within SGET contain polynomial and hyperbolic functions that depend on the material's length scale parameters (ratio between gradient and standard elastic moduli) and inclusion size \cite{ma2018inclusion}. Similarly to classical elasticity, the shape of inclusions define the structure of Eshelby-like tensors, while these tensors do not depend on the elastic properties of inclusions (see \cite{ma2018inclusion}). Therefore, similarly to classical self-consistent approach \cite{markov2000elementary, mura2013micromechanics}, one can use the averaged Eshelby-like tensors in \eqref{fin} that were obtained within SGET inclusion problems for isotropic matrix materials.

Contractions and inversions of the fourth-order and six-order tensors in \eqref{fin} can be performed in a simple manner by using its matrix representations and generalized Voigt notations \cite{auffray2019complete, yang2022verification}, or in a most straightforward way by using appropriate orthogonal bases that are well known for the fourth-order tensors and that were developed for the class of transversally isotropic six-order tensors in Ref. \cite{monchiet2013algebra}.

Note, that relations \eqref{eqv} can be alternatively derived considering equality conditions for the averaged stress $\pmb\tau$ and double stress $\pmb\mu$ inside a grain and corresponding macro-scale stresses. Although, in this case it will be not possible to define the relation for the second gradient of displacement \eqref{eqv}$_2$ in the case of Cauchy-type material of grain. Thus, it seems that the presented kinematic approach \eqref{qbc}-\eqref{eqv} will be more preferable within SGET.

Classical relations of the self-consistent method can be directly obtain from \eqref{eqv} neglecting second gradient condition \eqref{eqv}$_2$ and assuming that the effective medium is a Cauchy-type continuum so that Eshelby tensor $\langle\textbf S^{(4)}\rangle_{v}$ becomes classical.

\section{Solution for the Cauchy-type spherical grains}

Let us assume that the gradient moduli of crystallites  are very small ($\textbf G^{gr} \approx 0$) and has spherical shape. The first assumption is realistic and it bases on the known results of atomistic calculations for the gradient moduli of monocrystals, for which it was found that the the length scale parameters of SGET has sub-nanometer order \cite{shodja2018toupin,lazar2022mathematical}. The assumption about spherical shape of grains is just an approximation of the model that is widely used within classical micromechanics \cite{kroner1958berechnung,markov2000elementary}. For the considered case of Cauchy-type grains, the relation for the averaged second gradient of displacement \eqref{fin}$_2$ is reduced to:
\begin{equation}
\label{sph}
\left\langle\left(\textbf I^{(6)} - \langle\textbf S^{(6)}\rangle_v\right)^{-1}\right\rangle_{\Omega} = \textbf I^{(6)}
\end{equation}

In the left hand side of this relation we have isotropic identity tensor and Eshelby-like tensor that is also isotropic in the case of spherical inclusions \cite{ma2018inclusion}. Therefore, the procedure of averaging over orientations can be omitted and this relation can be immediately reduced to the following condition: 
\begin{equation}
\label{zs}
\langle\textbf S^{(6)}\rangle_v=0
\end{equation}

From the known representation of tensor $\langle\textbf S^{(6)}\rangle_v$ for spherical inclusions within SGET (see \cite{ma2018inclusion}) it follows that all effective gradient moduli should be zero $a_i=0$ to provide the fulfilment of obtained condition \eqref{zs}. Moreover, the fourth-order Eshelby tensor becomes classical in this case ($\langle\textbf S^{(4)}\rangle_v = \textbf S^{(4)}$, see \cite{ma2018inclusion,Ma2014}) and the first homogenization condition \eqref{fin}$_1$ reduces to those one of classical Kr{\"o}ner's method for the effective Lame constants $\lambda$ and $\mu$. Thus, for the case of low gradient moduli and spherical shape of grains the presented second-gradient self-consistent method provides classical solution for the effective polycrystalline media with the absence of strain gradient effects.

\section{Possible improvements of self-consistent estimates within SGET}
\label{ext}

The main next step that should be performed within the considered approach is the derivation of the solution for the grains of a spheroidal shape. In this case the averaged six-order Eshelby tensor in condition \eqref{fin}$_2$ will be non-isotropic and we can obtain the non-trivial system of seven coupled equations for determination of the effective gradient moduli and Lame constants of polycrystals. This system can be additionally simplified without much loss of accuracy assuming infinitely small values of individual gradient moduli of grains and using \eqref{sph} instead of \eqref{fin}$_2$. However, for the best of author's knowledge, up to date the closed form expressions for tensor $\langle\textbf S^{(6)}\rangle_v$ have  been established only for the spherical and circular inclusions within SGET \cite{ma2018inclusion}. Thus, we want to stress out the need of such studies for the more general shape of inclusions in the future work.

Also it is notable that Kr{\"o}ner's self-consistent method can be extended not only for the case of linear distribution of macro-scale strain field \eqref{qbc} but for the more general external conditions. Consideration of the $N$-th order polynomial distribution of the strain field together with additional mapping conditions for the averaged gradients of displacement up to ($N$+1)-th order (or gradients of strain field up to $N$-th order) will allow to derive the relations for the effective material constants of polycristals within high-grade elasticity theories in a similar manner that it is done within Mindlin Form I in the present contribution. 

There may also arise a question about possibility of introduction of additional mapping conditions between the micro- and macro- strain fields. Without dwelling on the energy-based conditions that were discussed, e.g. in Refs. \cite{Ganghoffer2021, Forest2011}, let us consider the possible additional kinematical measure evaluated as averaged strain field multiplied with position vector: $\langle\pmb\varepsilon^{gr}\otimes\textbf x\rangle_v$. Such quantity arises in the second-gradient self-consistent homogenization of composites  \cite{solyaev2022self} and its static counterpart is an averaged first static moment of stress $\langle\pmb\tau^{gr}\otimes\textbf x\rangle_v$ \cite{Ma2014,Ganghoffer2021}. Based on the extended Eshelby's equivalent inclusion method it was shown that under prescribed quadratic boundary conditions \eqref{qbc} the following relation is valid within SGET \cite{solyaev2022self}:
\begin{equation}
\label{ex}
\begin{aligned}
	\langle\pmb\varepsilon^{gr}\otimes\textbf x\rangle_v  
		&= \langle\tilde{\textbf B}\rangle_v
			\vdots \,\pmb\kappa^0
\end{aligned}
\end{equation}
where $\langle\tilde{\textbf B}\rangle_v$ is the six-order concentration tensor of special kind, which explicit representation within SGET was given in Ref. \cite{solyaev2022self}; for the case of Cauchy-type inclusion it was shown that this tensor is simply defined as $\langle\tilde{\textbf B}\rangle_v = \textbf I^{(6)}\cdot\langle\textbf x\otimes\textbf x\rangle_v$. Therefore, from \eqref{ex}  we obtain:
\begin{equation}
\label{ex1}
\begin{aligned}
	\langle\pmb\varepsilon^{gr}\otimes\textbf x\rangle_v  
		&= (\textbf I^{(6)}\cdot\langle\textbf x\otimes\textbf x\rangle_v) 
			\vdots \,\pmb\kappa^0 = \tfrac{1}{5}R^2 \pmb\kappa^0
\end{aligned}
\end{equation}
where we assume that the grain is spherical so that $\langle\textbf x\otimes\textbf x\rangle_v = \tfrac{1}{5}R^2 \textbf I^{(2)}$.
 
Thus, the averaging procedure and mapping to the corresponding macro-scale field  become trivial for this measure:
\begin{equation}
\label{ex2}
\begin{aligned}
	\langle\pmb\varepsilon^{gr}\otimes\textbf x\rangle_{v,\Omega}
	\equiv 
	\langle\pmb\varepsilon(\textbf x)\otimes\textbf x\rangle_{v,\Omega}
	=\langle\pmb\varepsilon^0\otimes\textbf x\rangle_{v,\Omega}
	+\pmb\kappa^0\langle\textbf x\otimes\textbf x\rangle_{v,\Omega}
		= \tfrac{1}{5}R^2 \pmb\kappa^0
\end{aligned}
\end{equation}
where we take into account that the origin of coordinates is placed at the center of spherical grain. Similar trivial result can be also obtained for more complex shape of randomly oriented grains so that we can state that within the proposed second-gradient homogenization approach it is enough to use only two mapping conditions \eqref{cond}.


\section{Conclusions}
\label{con}

In this note we suggest a framework for determination of the effective second-gradient Mindlin elastic properties of polycrystalline materials. Kr{\"o}ner's self-consistent scheme is extended assuming that the equivalent effective medium can be described by Mindlin-Toupin SGET. Additional micro-macro scale mapping condition for the second gradient of displacement is introduced. 
Considered approach predicts absence of gradient effects for the case of randomly oriented spherical grains. However, more general shape of grains (e.g. spheroidal) could be the source of non-local gradient effects in the material microstructure. A set of non-linear relations given by Eqs. \eqref{fin} can be used for determination of the effective standard and gradient moduli of polycrystalline materials. Corresponding explicit representation for the six-order Eshelby-like tensors should derived within SGET and used in such calculations. 

Importance and practical application of the obtained results can be related to the description of size-effects in fracture of quasi-brittle polycristaline materials by using estimated values of gradient moduli \cite{vasiliev2021new,placidi2022hemi} and also for the micromechanical validation of the known phenomenological simplified constitutive models of SGET \cite{polizzotto2017hierarchy,polizzotto2018anisotropy}.

\section*{References}
\renewcommand{\bibsection}{}
\bibliography{refs.bib}

\begin{thebibliography}{10}

\bibitem{hershey1954elasticity}
AV0059 Hershey.
\newblock The elasticity of an isotropic aggregate of anisotropic cubic
  crystals.
\newblock 1954.

\bibitem{kroner1958berechnung}
Ekkehart Kr{\"o}ner.
\newblock Berechnung der elastischen konstanten des vielkristalls aus den
  konstanten des einkristalls.
\newblock {\em Zeitschrift f{\"u}r Physik}, 151(4):504--518, 1958.

\bibitem{kube2015ultrasonic}
Christopher~M Kube and Joseph~A Turner.
\newblock Ultrasonic attenuation in polycrystals using a self-consistent
  approach.
\newblock {\em Wave Motion}, 57:182--193, 2015.

\bibitem{kneer1965berechnung}
G~Kneer.
\newblock {\"U}ber die berechnung der elastizit{\"a}tsmoduln vielkristalliner
  aggregate mit textur.
\newblock {\em Physica status solidi (b)}, 9(3):825--838, 1965.

\bibitem{kanaun1980poisson}
SK~Kanaun.
\newblock The poisson set of cracks in an elastic continuous medium.
\newblock {\em Journal of Applied Mathematics and Mechanics}, 44(6):808--815,
  1980.

\bibitem{chen2022multiscale}
Qiang Chen, Fengyuan Zhao, Jinhao Jia, Changjun Zhu, Shuxin Bai, and Yicong Ye.
\newblock Multiscale simulation of elastic response and residual stress for
  ceramic particle reinforced composites.
\newblock {\em Ceramics International}, 48(2):2431--2440, 2022.

\bibitem{kroner1961plastischen}
E~Kr{\"o}ner.
\newblock Zur plastischen verformung des vielkristalls.
\newblock {\em Acta metallurgica}, 9(2):155--161, 1961.

\bibitem{lubarda1997new}
Vlado~A Lubarda.
\newblock New estimates of the third-order elastic constants for isotropic
  aggregates of cubic crystals.
\newblock {\em Journal of the Mechanics and Physics of Solids}, 45(4):471--490,
  1997.

\bibitem{sevostianov2019kroner}
Igor Sevostianov and Marat~R Talipov.
\newblock Kr{\"o}ner method for thermal or electrical conductivity of
  polycrystals and other aggregates of anisotropic particles.
\newblock {\em International Journal of Engineering Science}, 136:67--77, 2019.

\bibitem{Mindlin1964}
R.~D. Mindlin.
\newblock {Micro-structure in linear elasticity}.
\newblock {\em Archive for Rational Mechanics and Analysis}, 16(1):51--78,
  1964.

\bibitem{Toupin1964}
R.~A. Toupin.
\newblock {Theories of elasticity with couple-stress}.
\newblock {\em Archive for Rational Mechanics and Analysis}, 17(2):85--112,
  1964.

\bibitem{dell2009generalized}
Francesco Dell'Isola, Giulio Sciarra, and Stefano Vidoli.
\newblock Generalized hooke's law for isotropic second gradient materials.
\newblock {\em Proceedings of the Royal Society A: Mathematical, Physical and
  Engineering Sciences}, 465(2107):2177--2196, 2009.

\bibitem{Askes2015}
H.~Askes and L.~Susmel.
\newblock {Understanding cracked materials: Is Linear elastic fracture
  mechanics obsolete?}
\newblock {\em Fatigue and Fracture of Engineering Materials and Structures},
  38(2):154--160, 2015.

\bibitem{vasiliev2021new}
Valeriy Vasiliev, Sergey Lurie, and Yury Solyaev.
\newblock New approach to failure of pre-cracked brittle materials based on
  regularized solutions of strain gradient elasticity.
\newblock {\em Engineering Fracture Mechanics}, page 108080, 2021.

\bibitem{placidi2022hemi}
Luca Placidi, Emilio Barchiesi, Francesco Dell'Isola, Valerii Maksimov, Anil
  Misra, Nasrin Rezaei, Angelo Scrofani, and Dmitry Timofeev.
\newblock On a hemi-variational formulation for a 2d elasto-plastic-damage
  strain gradient solid with granular microstructure.
\newblock {\em Mathematics in Engineering}, 5:1--24, 2022.

\bibitem{makvandi2021strain}
Resam Makvandi, Bilen~Emek Abali, Sascha Eisentr{\"a}ger, and Daniel Juhre.
\newblock A strain gradient enhanced model for the phase-field approach to
  fracture.
\newblock {\em PAMM}, 21(1):e202100195, 2021.

\bibitem{nejadsadeghi2020role}
Nima Nejadsadeghi and Anil Misra.
\newblock Role of higher-order inertia in modulating elastic wave dispersion in
  materials with granular microstructure.
\newblock {\em International Journal of Mechanical Sciences}, 185:105867, 2020.

\bibitem{dell2019pantographic}
Francesco Dell’Isola, Pierre Seppecher, Jean~Jacques Alibert, Tomasz
  Lekszycki, Roman Grygoruk, Marek Pawlikowski, David Steigmann, Ivan Giorgio,
  Ugo Andreaus, Emilio Turco, et~al.
\newblock Pantographic metamaterials: an example of mathematically driven
  design and of its technological challenges.
\newblock {\em Continuum Mechanics and Thermodynamics}, 31:851--884, 2019.

\bibitem{bacca2013anisotropic}
Mattia Bacca, Francesco Dal~Corso, Daniele Veber, and Davide Bigoni.
\newblock Anisotropic effective higher-order response of heterogeneous cauchy
  elastic materials.
\newblock {\em Mechanics Research Communications}, 54:63--71, 2013.

\bibitem{aydin2022strain}
Gokhan Aydin, M~Erden Yildizdag, and Bilen~Emek Abali.
\newblock Strain-gradient modeling and computation of 3-d printed metamaterials
  for verifying constitutive parameters determined by asymptotic
  homogenization.
\newblock In {\em Theoretical Analyses, Computations, and Experiments of
  Multiscale Materials: A Tribute to Francesco dell’Isola}, pages 343--357.
  Springer, 2022.

\bibitem{solyaev2022self}
Yury Solyaev.
\newblock Self-consistent assessments for the effective properties of two-phase
  composites within strain gradient elasticity.
\newblock {\em Mechanics of Materials}, 169:104321, 2022.

\bibitem{yang2022verification}
Hua Yang, B~Emek Abali, Wolfgang~H M{\"u}ller, Salma Barboura, and Jia Li.
\newblock Verification of asymptotic homogenization method developed for
  periodic architected materials in strain gradient continuum.
\newblock {\em International Journal of Solids and Structures}, 238:111386,
  2022.

\bibitem{rahali2015homogenization}
Y~Rahali, I~Giorgio, JF~Ganghoffer, and Francesco dell'Isola.
\newblock Homogenization {\`a} la piola produces second gradient continuum
  models for linear pantographic lattices.
\newblock {\em International Journal of Engineering Science}, 97:148--172,
  2015.

\bibitem{lahbazi2022size}
Ahmed Lahbazi, Ibrahim Goda, and Jean-Fran{\c{c}}ois Ganghoffer.
\newblock Size-independent strain gradient effective models based on
  homogenization methods: applications to 3d composite materials, pantograph
  and thin walled lattices.
\newblock {\em Composite Structures}, 284:115065, 2022.

\bibitem{solyaev2022effective}
Y~Solyaev.
\newblock Effective length scale parameters of the fiber-reinforced composites.
\newblock {\em Lobachevskii Journal of Mathematics}, 43(7):1993--2002, 2022.

\bibitem{barchiesi2021granular}
Emilio Barchiesi, Anil Misra, Luca Placidi, and Emilio Turco.
\newblock Granular micromechanics-based identification of isotropic strain
  gradient parameters for elastic geometrically nonlinear deformations.
\newblock {\em ZAMM-Journal of Applied Mathematics and Mechanics/Zeitschrift
  f{\"u}r Angewandte Mathematik und Mechanik}, 101(11):e202100059, 2021.

\bibitem{Ma2014}
H.~M. Ma and Xin~L. Gao.
\newblock A new homogenization method based on a simplified strain gradient
  elasticity theory.
\newblock {\em Acta Mechanica}, 225(4-5):1075--1091, 2014.

\bibitem{ma2018inclusion}
Hansong Ma, Gengkai Hu, Yueguang Wei, and Lihong Liang.
\newblock Inclusion problem in second gradient elasticity.
\newblock {\em International Journal of Engineering Science}, 132:60--78, 2018.

\bibitem{polizzotto2016note}
Castrenze Polizzotto.
\newblock A note on the higher order strain and stress tensors within
  deformation gradient elasticity theories: Physical interpretations and
  comparisons.
\newblock {\em International Journal of Solids and Structures}, 90:116--121,
  2016.

\bibitem{auffray2019complete}
Nicolas Auffray, Qi-Chang He, and Hung Le~Quang.
\newblock Complete symmetry classification and compact matrix representations
  for 3d strain gradient elasticity.
\newblock {\em International Journal of Solids and Structures}, 159:197--210,
  2019.

\bibitem{po2019green}
Giacomo Po, Nikhil~Chandra Admal, and Markus Lazar.
\newblock The green tensor of mindlin’s anisotropic first strain gradient
  elasticity.
\newblock {\em Materials Theory}, 3(1):1--16, 2019.

\bibitem{polizzotto2018anisotropy}
Castrenze Polizzotto.
\newblock Anisotropy in strain gradient elasticity: Simplified models with
  different forms of internal length and moduli tensors.
\newblock {\em European Journal of Mechanics-A/Solids}, 71:51--63, 2018.

\bibitem{markov2000elementary}
Konstantin~Z Markov.
\newblock Elementary micromechanics of heterogeneous media, 2000.

\bibitem{Forest2011}
S.~Forest and D.~K. Trinh.
\newblock {Generalized continua and non-homogeneous boundary conditions in
  homogenisation methods}.
\newblock {\em ZAMM Zeitschrift fur Angewandte Mathematik und Mechanik},
  91(2):90--109, 2011.

\bibitem{monchiet2020strain}
Vincent Monchiet, Nicolas Auffray, and Julien Yvonnet.
\newblock Strain-gradient homogenization: a bridge between the asymptotic
  expansion and quadratic boundary condition methods.
\newblock {\em Mechanics of Materials}, 143:103309, 2020.

\bibitem{mura2013micromechanics}
Toshio Mura.
\newblock {\em Micromechanics of defects in solids}.
\newblock Springer Science \& Business Media, 1982.

\bibitem{monchiet2013algebra}
Vincent Monchiet and Guy Bonnet.
\newblock Algebra of transversely isotropic sixth order tensors and solution to
  higher order inhomogeneity problems.
\newblock {\em Journal of Elasticity}, 110:159--183, 2013.

\bibitem{shodja2018toupin}
Hossein~M Shodja, Hashem Moosavian, and Farzaneh Ojaghnezhad.
\newblock Toupin--mindlin first strain gradient theory revisited for cubic
  crystals of hexoctahedral class: analytical expression of the material
  parameters in terms of the atomic force constants and evaluation via ab
  initio dft.
\newblock {\em Mechanics of Materials}, 123:19--29, 2018.

\bibitem{lazar2022mathematical}
Markus Lazar, Eleni Agiasofitou, and Thomas B{\"o}hlke.
\newblock Mathematical modeling of the elastic properties of cubic crystals at
  small scales based on the toupin--mindlin anisotropic first strain gradient
  elasticity.
\newblock {\em Continuum Mechanics and Thermodynamics}, 34(1):107--136, 2022.

\bibitem{Ganghoffer2021}
J.~F. Ganghoffer and H.~Reda.
\newblock A variational approach of homogenization of heterogeneous materials
  towards second gradient continua.
\newblock {\em Mechanics of Materials}, 158(December 2020):103743, 2021.

\bibitem{polizzotto2017hierarchy}
Castrenze Polizzotto.
\newblock A hierarchy of simplified constitutive models within isotropic strain
  gradient elasticity.
\newblock {\em European Journal of Mechanics-A/Solids}, 61:92--109, 2017.

\end{thebibliography}

\end{document}